\documentclass
{mn2e}
\usepackage{epsfig}
\usepackage{amsmath, amssymb,bm}

\def \be{\begin{equation}}
\def \ee{\end{equation}}
\def \rsun{\rm R_{\odot}}
\def \msun{\rm M_{\odot}}

\begin{document}
%%%%%%%%%%%%%%%%%%%%%%%%%%%%%%%%%%%%%%%%%%%%%%%%%%%%%%%%%%%%%%%%%%%%%%%%%%%%%%
%% Title Details and Page Header                                            %%
%%%%%%%%%%%%%%%%%%%%%%%%%%%%%%%%%%%%%%%%%%%%%%%%%%%%%%%%%%%%%%%%%%%%%%%%%%%%%%
\title[GSN 069 -- A Tidal Disruption Near--Miss]{GSN 069 -- A Tidal Disruption Near--Miss}

\author[Andrew King] 
{\parbox{5in}{Andrew King$^{1, 2, 3}$ 
}
\vspace{0.1in} \\ $^1$ School of Physics \& Astronomy, University
of Leicester, Leicester LE1 7RH UK\\ 
$^2$ Astronomical Institute Anton Pannekoek, University of Amsterdam, Science Park 904, NL-1098 XH Amsterdam, The Netherlands \\
$^{3}$ Leiden Observatory, Leiden University, Niels Bohrweg 2, NL-2333 CA Leiden, The Netherlands}

\maketitle

\begin{abstract}
I suggest that the quasiperiodic ultrasoft X--ray eruptions recently 
observed from the galaxy GSN 069 may result from
accretion from a low--mass white dwarf in a highly eccentric orbit 
about its central black hole. At $0.21\msun$, this star was probably
the core of a captured red giant. Such events should occur in significant numbers as less extreme outcomes of whatever process leads to tidal 
disruption events. I show that gravitational radiation losses can 
drive the observed mass transfer rate, and that the precession of the
white dwarf orbit may be detectable in X--rays as a superorbital 
quasiperiod $P_{\rm super} \simeq 2\,{\rm d}$. 
The very short lifetime of the current event, and the
likelihood that similar ones involving more massive stars
would be less observable,
together suggest that stars may 
transfer mass 
% drive accretion on 
to the low--mass SMBH in this and similar galaxies 
at a total rate potentially 
making a significant contribution to their masses. A similar 
or even much greater inflow rate would be unobservable in most galaxies. I 
discuss the 
implications for SMBH mass growth. 
\end{abstract}


\footnotetext[1]{E-mail: ark@astro.le.ac.uk}
\section{Introduction}
\label{intro}
Miniutti et al. (2019) have recently discovered large--amplitude (factors $\sim 
100$) quasi--periodic
X--ray eruptions from the low--mass black hole ($M_1 \sim 4\times 
10^5\msun$) galaxy nucleus  GSN 069. 
These each last a little more than 1 hr, with a characteristic recurrence 
time $\simeq 9$\,hr. The emission has 
an ultrasoft blackbody spectrum with peak temperature and luminosity $T 
\simeq10^6\,{\rm K}, L\simeq 5\times 
10^{42}\, {\rm erg\, s^{-1}}$. These imply a blackbody radius $R_{bb} =
9\times 10^{10}$\, cm, slightly larger than the gravitational radius 
$R_g = GM_1/c^2 = 6\times 10^{10}\,{\rm cm}$ of the black hole.

The very large large amplitudes and short timescales of the eruptions are 
difficult to explain except as mass transfer events.
The quasiperiodic repetitions suggest that mass overflowing 
from a star in an elliptical 9--hour orbit about the 
black hole triggers powerful instabilities in the accretion disc at each
pericentre passage. Hysteresis effects probably account for the departure 
from 
strict periodicity in the X--ray emission, as in stellar--mass systems of this 
type. 
\section{Mass Transfer}
Adopting this view, we have significant constraints on the orbiting star. The 
orbital semimajor axis is 
\be
a = 1\times 10^{13}m_{5.6}^{1/3}P_9^{2/3}\,{\rm cm}
\label{a}
\ee
where $m_{5.6}$ is the black hole mass $M_1$ in units of $4\times 
10^5\msun$ and $P_9$ is the orbital period in units of 9 hr.
The tidal lobe
of the orbiting star, of mass $M_2$, is
\be
R_{\rm lobe} \simeq 0.46(M_2/M_1)^{1/3}a(1-e) \simeq 6.2\times 
10^{10}m_2^{1/3}(1-e)\,{\rm cm},
\label{lobe}
\ee 
where $m_2 = M_2/\msun$, $e$ is the eccentricity, and I have adopted the 
prescription of Sepinsky et al. (2007) for tidal overflow in eccentric binaries 
(see also Dosopolou \& Kalogera, 2016a, b).  Since the star's radius $R_2 
= r_2\rsun$ must equal $R_{\rm lobe}$ at pericentre, it must currently 
obey the constraint
\be
r_2 = 0.89m_2^{1/3}(1-e)
\label{lobe}
\ee
(note that this is {\it not} the mass--radius relation of the star, but
simply requires that that relation must give values of $r_2, m_2$ obeying
(\ref{lobe}) at the present epoch).

The gas lost from the orbiting star at pericentre passage circularizes at  
radius 
\be
R_{\rm circ}\sim a(1-e) \simeq 10^{13}(1-e)\,{\rm cm}. 
\label{circ}
\ee
resulting in the formation of an accretion disc of outer radius 
$R_d \sim R_{\rm circ}$.

I now ask if this kind of binary system can generate the very
large mass transfer rates required to explain the accretion luminosity.
I assume that the observed rate given by the repeated outbursts is
representative of the evolutionary mean, and justify this assumption later.

For a typical black 
hole accretion efficiency of 10\% the X--ray eruptions require a
mass accretion rate
$\sim 5\times 10^{22}\,{\rm g\,s^{-1}}$ at peak. 
Averaging these over the full 9--hour cycle gives a 
mean mass transfer rate 
\be
-\dot M_2 \simeq 10^{-4}\,\msun\,{\rm yr}^{-1},
\label{mt}
\ee
and so a mass--transfer timescale 
\be
t_{\dot M} \sim -M_2/\dot M_2 \sim 10^4m_2\,{\rm yr.}
\label{MT}
\ee
In general there are only two ways to drive significant mass transfer rates 
in a binary system: either the mass--losing star which fills its tidal lobe at 
pericentre must expand on the timescale $t_{\dot M}$, or the binary must
lose orbital angular momentum on this timescale. 

The first possibility is very 
unlikely: no known star has nuclear or thermal timescales as short as 
(\ref{MT}), and dynamical timescale mass transfer implies a timescale 
$t_{\dot M}$ of only a few orbits, probably resulting in a common
envelope, contrary to observation. So the system must instead lose
orbital angular momentum on the timescale $t_{\dot M}$. 

The only likely
mechanism for this is gravitational radiation (GR), which is potentially
very efficient here because of the short orbital period and high total mass.
The system then resembles a drastically speeded--up and eccentric 
version of short--period cataclysmic variable (CV) 
evolution. This is a long--studied area,
(e.g. Faulkner, 1971; Pacz\'ynski \& Sienkiewicz, 1981; see King, 1988 for 
a review). Hameury et al., (1994)
Dai \& Blandford (2013) and Linial \& Sari (2017) discuss low--mass
stars in circular orbits around SMBH.

For an eccentric orbit, the quadrupole GR loss rate is given by 
\be
\frac{\dot J}{J} _{\rm GR} = -\frac{32}{5}\frac{G^3}{c^5}
\frac{M_1M_2M}{a^4}f(e),
\label{GR}
\ee
where $J$ is the orbital angular momentum, $M = M_1 + M_2$
the (constant) total mass, and
\be 
f(e) = {1 + {73\over 24}e^2 + {37\over 96}e^4\over (1 - e^2)^{7/2}}
\label{f}
\ee
(Peters \& Mathews, 1963). This shrinks the semimajor axis $a$ 
while reducing the eccentricity more rapidly. These quantities
are related by
\be 
a = \frac{c_0 e^{12/19}}{1 - e^2}
\left(1 + \frac{121}{304}e^2\right)^{870/2299}
\label{ae}
\ee
(Peters, 1964), where $c_0$ is a constant set by the initial value of $a$. We see that for extreme eccentricities $e \sim 1$ (i.e. $1 - e << 1$)
we have
\be
a \propto \frac{1}{1 - e^2} \sim \frac{1}{2(1 - e)}
\label{e}
\ee
so we set
\be
a = \frac{1-e_0}{1-e}a_0
\label{a0}
\ee
where 
$a_0 = 1\times 10^{13} M_{5.6}^{1/3}P_9^{2/3}\,{\rm cm}$ and $e_0 \simeq 1$
are the current semimajor axis and eccentricity.

This implies that the
pericentre separation
\be
a(1 - e) \simeq a_0(1 - e_0)
\ee
remains almost constant when $e \sim 1$
-- this is reasonable, since the GR emission
is effectively confined to a point interaction at pericentre. 
From 
(\ref{e}) the orbital angular momentum
\be
J = M_1M_2\left(\frac{Ga}{M}\right)^{1/2}(1 - e^2)^{1/2} 
\simeq M_1M_2\left(\frac{Ga_0}{M}\right)^{1/2}(1-e_0^2)^{1/2}
\ee
simply varies as $J \propto M_1M_2$. Logarithmic differentiation now 
gives
\be 
\frac{\dot J}{J} = \frac{\dot M_1}{M_1} + \frac{\dot M_2}{M_2}
 = \frac{\dot M_2}{M_2}\left(1 - \frac{M_2}{M_1}\right) \simeq 
\frac{\dot M_2}{M_2}
\ee
so that the current GR--driven mass transfer rate is
\be
-\dot M_2\simeq
1\times 10^{-7}m_{5.6}^{2/3}P_9^{-8/3}
\frac{m_2^2}{(1-e_0)^{7/2}}
\,\msun\,{\rm yr}^{-1},
\label{grmt}
\ee
where $e$ is set $=1$ except in factors $(1-e)$.
 
The theoretical rate (\ref{grmt}) gives the evolutionary mean mass transfer, 
evaluated over the time $t_{\rm lobe}$ the tidal lobe takes to move
through one density scaleheight of the star. This is
typically about $10^{-4}$.  
%% about $10^{-4{R)2$ near
the inner Lagrange 
point (Ritter, 1988) so here
\be
t_{\rm lobe} \sim 10^{-4}\frac{R_2}{\dot R_2} 
\sim 10^{-4}\frac{M_2}{|\dot M_2|} \sim 0.1\,{\rm yr}.
\ee
Normally $t_{\rm lobe}$ is far longer than the observing timescale, but 
here (uniquely) this is reversed because the mass transfer timescale is very 
short. The currently observed accretion rate is a good indicator of the 
long--term evolutionary mean, as asserted above.

\section{The Orbiting Star}
The work of the last Section gives two constraints 
(eqns (\ref{lobe}) and (\ref{grmt})), which 
simultaneously fix the 
mass of the orbiting star and the eccentricity $e$. For a plausible
identification, these values must be consistent with a physically
reasonable mass--radius relation. The extremely short mass--transfer 
timescale $t_{\dot M} \sim 10^4\,{\rm yr}$ already tells us that
this must either be set by the adiabatic reaction of a non--degenerate star 
to adiabatic 
%% rapid
mass loss (cf Dai et al., 2013), or correspond to a degenerate
star (e.g. a white dwarf).

First,
to provide the deduced mass transfer rate $\sim 10^{-4}\msun\,{\rm 
yr}^{-1}$, eqn (\ref{grmt}) requires
\be
\frac{m_2^2}{(1 - e_0)^{7/2}} = 10^3,
\label{rmt}
\ee 
or 
\be
1 - e_0 \simeq 0.14m_2^{4/7}.
\label{ee}
\ee
Substituting this into eqn (\ref{lobe}) gives
\be
R_2 = r_2\rsun = 8.7\times 10^9m_2^{0.91}\,{\rm cm}.
\label{mr}
\ee
This radius is so small for any reasonable stellar mass $m_2$ that the only 
possibility is a low--mass white dwarf, whose mass--radius relation we can 
take as 
\be 
R_2 \simeq 1\times 10^9(m_2/0.5)^{-1/3}\, {\rm cm}
\label{mr2}.
\ee
We see that eqns (\ref{mr}, \ref{mr2}) are compatible for 
\be
M_2 = 0.21\msun,
\label{m2}
\ee
while from (\ref{ee}) we find a self--consistently large current eccentricity
\be
e_0 = 0.94.
\ee

The future evolution of the system is straightforward: as the white dwarf 
expands on mass loss and the eccentricity decreases, the mass transfer
rate will drop sharply (typically as $\sim M_2^5$, cf the similar evolution
of very short--period CVs, e.g. King, 1988). The system will transfer mass
at ever--slowing rates almost indefinitely.

\section{Origin}

An important consequence of the low value of $M_2$ is that
the current mass transfer timescale is very short, i.e 
$t_{\dot M} \sim M_2/(-\dot M_2) \sim 2000\,{\rm yr}$. That
we are nevertheless able to observe such a brief event
means that the rate of similar events must be very high. Together with
the low SMBH mass, these facts strongly favour some kind of 
tidal capture event as the basic origin of this kind of system.
It is also clear that the current 
mass $M_2 \simeq 0.21\msun$ of the orbiting white dwarf is too
low to be the straightforward outcome of single--star evolution. There
are two obvious possibilities. 

(a) The white dwarf began mass transfer with a `normal' mass $M_2
\simeq 0.6\msun$. It is easy to show (e.g. King, 1988) that with 
mass--radius relation $R_2 \propto M_2$ 
% $R_2 \propto M_2^{-1/3}$
the orbital period goes as 
$P \propto M_2^{-1}$, so the original period must have been only
$\simeq 3\,{\rm hr}$. 

(b) The white dwarf was originally the core of a red giant. If it is
still close to its mass at that epoch, the red giant would have had a
radius $\sim 12\rsun$. If instead the current white dwarf
has already transferred a large fraction 
of its original mass, the giant would have been much larger. In all
cases the red giant envelope could have had a significant mass.

Case (b) is considerably more likely, as it allows the interpretation
that the current system is the survivor of some kind of tidal capture of
a red giant (whereas Case (a) requires an `aim' of implausible accuracy).
Case (b) could have been triggered by a full tidal disruption event (TDE)
involving explosive mass transfer, but a near--miss event in which the giant 
was captured into an orbit where it eventually
lost mass only at pericentre (as the white dwarf does now) is more 
probable. The conditions for a TDE are extremely restrictive, so such
near--miss events must more far more common. Mass
transfer would have stopped for a time once the giant lost its envelope. 
The binary separation at this point would have
been noticeably wider
($a(1-e) \sim 7\times 10^{13}\,{\rm cm}$
for the minimum giant radius of $12\rsun$), but 
gravitational wave emission 
would have made the white dwarf core fill its tidal lobe on a relatively 
short timescale, because of the high eccentricity.

\section{Implications for SMBH Feeding}

I have argued above that 
tidal near--miss events like the one studied here must be quite  
common, suggesting that similar events with different infalling stars
should occur also. But it seems likely that the particular event
studied here was unusually favoured for observation, as one might
expect, given that it is currently fairly unique. The favoritism arises
because the very small stellar radius, and hence very high mass density, 
means that mass transfer starts only
at a rather small pericentre distance, where gravitational radiation can
drive very rapid mass transfer. It is a well--known result of Roche geometry 
(e.g King, 1988) that the mean mass density $\bar\rho$ 
of a lobe--filling star goes as $P^{-2}$, where $P$ is the orbital period.
This is aided still more in the present case 
because of the high eccentricity. More massive and therefore more 
extended infalling stars fill their Roche lobes at wider separations, 
corresponding to much longer orbital periods $P$. From (\ref{grmt}) we 
see that the $P^{-8/3}$ dependence of the GR--driven mass transfer 
rate is likely to outweigh the $m_2^2$ effect of an increased stellar mass.

This suggests that the present event may be only the most observable 
component of a considerably larger infall rate to the SMBH.
A similar -- or even far greater -- stellar infall rate in other 
galaxies would not be observable at all if the central SMBH is massive 
enough that the infalling stars are swallowed by the black hole before they
fill their tidal lobes, which happens if $M \ga 10^7\msun$ (Kesden, 
2012).
At low 
redshift we know  from the Soltan (1982)  relation that a mechanism like
this cannot be supplying most of the total SMBH mass. But it seems 
possible 
that it could be a significant contributor for smaller SMBH, and at higher
redshift.

\section{The Light Curve}

The eruptions characterizing the X--ray light curve of GSN~069 have far
shorter timescales than are likely for the usual diffusive viscous transport in 
accretion discs. They resemble the light curve of of GRS~1915+105 (Belloni 
et al. 1997), which shows evidence for the viscous refilling of a disc depleted 
by flares. This kind of behaviour is modelled by King et al. (2004), who
suggest that local dynamo processes can affect the evolution of an accretion 
disc by driving angular momentum loss in the form of an outflow (a wind or 
jet). The waiting timescale for such eruptions to occur is much shorter
if the disc is thick ($H \sim R$) as it is triggered by the chance alignment
of local magnetic fields anchored in adjacent disc annuli, which has a 
timescale $\sim 2^{R/H}t_{\rm dyn}$, where $t_{\rm dyn}$ is the 
local disc dynamical timescale $(R_d^3/GM)^{1/2}$.
We will see below that there is reason to expect a thick
outer disc in this system. 

The form of the X--ray light curve must also be strongly affected because
the pericentre separation $p= (a(1-e_0) \simeq 6\times 10^{11}\,{\rm 
cm}$ is of order only $15R_g$. The standard formula
\be
\Delta \phi \simeq \frac{6\pi GM}{c^2a(1-e)}
\ee
for pericentre advance now gives
\be
\frac{\Delta\phi}{2\pi} \simeq \frac{3R_g}{p} \simeq \frac{1}{5},
\ee
so pericentre precesses one full revolution roughly every 5 orbits. If the
inclination of the orbital plane to the line of sight is high enough, this
may be detectable as a superorbital quasiperiod $P_{\rm super} \sim 2\,
{\rm d}$ in X--rays.

\section{Conclusions}

I have shown that the 9~hr quasiperiodic X--ray eruptions from GSN~069
could result from mass overflow at pericentre from an orbiting low--mass
star in a very eccentric orbit. I have argued the systems like this would result
from near--miss tidal disruption events, which should be considerably
more common than 
genuine TDEs. Similar events involving more massive stars would be less 
observable. In combination with the very short lifetime of the current event,
this suggests that stars fall close to the low--mass SMBH in this galaxy at
a rate $\ga 10^{-4}\msun\, {\rm yr}^{-1}$. A similar or even much 
greater inflow rate could have a major effect in growing SMBH masses,
either at high redshift, or in growing low--mass SMBH at low redshift, but
would be otherwise essentially unobservable.

This suggests several possible lines of future research.
We have seen that the mass 
transfer rate from any individual star falls very quickly below its initial 
value. Accordingly the SMBH might on average be accreting from several 
of them at low rates simultaneously. Their orbital planes are presumably 
uncorrelated, making the outer disc thick, and so favouring the 
dynamo--driven outbursts discussed above. 
Numerical simulations might check this picture, and see if
the sudden periodic injections of mass when the star is
at pericentre can trigger the eruptions.

Further X--ray observations of GSN 069 could potentially offer much
more insight into this system, particularly if the coverage is extensive
enough to offer the chance of detecting the predicted superorbital 
modulation $P_{\rm super} \simeq 2\,{\rm d}$. 
It is also clearly worthwhile checking other galaxies known to 
have low--mass SMBHs for similar quasiperiodic eruptions.

A final point concerns the nature of the orbiting star: if this is the fully--
stripped core of a red giant, the accreted material should be helium--rich. 
But it is possible that some of the envelope hydrogen may remain on the
surface. At present this question appears observationally intractable.

% Lense--Thirring, capture of a BINARY, 
\section*{Acknowledgments}

I thank Phil Uttley, Adam Ingram, Rhanna Starling and Andrew Blain for stimulating discussions. I am very grateful to the referee for a perceptive 
report.
%\begin{thebibliography}{}
\section*{REFERENCES}
Belloni T., Mendez M., King A. R., van der Klis M., van Paradijs J., 1997,
ApJ, 479, L145
\\
Dai, L. \& Blandford, R.D., 2013, MNRAS 434, 2948
\\
Dai, L., Blandford, R.D., \& Eggleton, P.P., 2013, MNRAS 434, 2930
\\ 
Dosopoulou, F., \& Kalogera, V., 2016a, ApJ 825, 70D
\\
Dosopoulou, F., \& Kalogera, V., 2016b, ApJ 825, 71D
\\
Faulkner, J., 1971, ApJL 170L, 99F
\\
Hameury, J.M., King, A.R., Lasota, J.P., Auvergne, M., 1994, A\&A 292, 404
\\
Kesden, M., 2012, Phys Rev D,85, 024037
\\
King, A.R., 1988, QJRAS 29,1
\\
King, A.R., Pringle, J.E., West, R.G., et al., 2004, MNRAS 348, 111
\\
Linial, I., \& Sari, R., 2017, MNRAS 469, 2441L 
\\
Miniutti, G., et al., 2019, Nature 573, 381
\\
Paczy\'nski, B.  \& Sienkiewicz, R., 1981, ApJLett 248, L27
\\
Peters, P.C., 1964, Phys Rev. 136, B1224
\\
Peters, P.C. \& Mathews, J., 1963, Phys Rev 131, 435
\\
Ritter, H., 1988, A\&A 202, 93
\\
Sepinsky, J. F., Willems, B., Kalogera, V., \& Rasio, F. A., 2007, ApJ
667, 1170
\\
Soltan, A, 1982, MNRAS 200, 115
%\end{thebibliography}{}

\end{document}